\documentclass[conference,a4paper]{IEEEtran}

\usepackage{amsmath,amsfonts,amssymb}
\usepackage{graphicx,color}
\usepackage{epstopdf}
\usepackage{tabls,stackrel,cite}
\usepackage{braket}
\usepackage{bm,notation}
\usepackage{balance}

\newcommand{\refE}[1]   {(\ref{eqn:#1})}

\IEEEoverridecommandlockouts

\begin{document}
\sloppy

\title{Multiple Quantum Hypothesis Testing Expressions\\
and Classical-Quantum Channel Converse Bounds}

\author{\authorblockN{Gonzalo Vazquez-Vilar}
\authorblockA{Universidad Carlos III de Madrid, Spain
%\\               Gregorio Mara\~n\'on Health Research Institute
}
\authorblockA{Email: gvazquez@ieee.org}
\thanks{G. Vazquez-Vilar is also with the Gregorio Mara\~n\'on Health Research Institute, Madrid, Spain. This work has been funded in part by the Spanish Ministry of Economy and Competitiveness under Grants FPDI-2013-18602, \mbox{TEC2013-41718-R}, and TEC2015-69648-REDC.}}

%% Create the title:
\maketitle

\begin{abstract}
Alternative exact expressions are derived for the minimum error probability of a hypothesis test discriminating among $M$ quantum states. The first expression corresponds to the error probability of a binary hypothesis test with certain parameters; the second involves the optimization of a given information-spectrum measure. Particularized in the classical-quantum channel coding setting, this characterization implies the tightness of two existing converse bounds; one derived by Matthews and Wehner using hypothesis-testing, and one obtained by Hayashi and Nagaoka via an information-spectrum approach.
\end{abstract}

\section{Introduction}\label{sec:intro}
%%%%%%%%%%%%%%%%%%%%%%%%%%%%%%%%%%%%%%%%%%%%%%%%%%%%%%%%%%%%%%%%%%%%%%%%%%

Optimal discrimination among multiple quantum states \mbox{--quantum hypothesis testing--} is at the core of several information processing tasks involving quantum-mechanical systems. 
%For example, quantum hypothesis testing yields precise fundamental limits on the detectability of optical signals~\cite{helstrom1967detection}. Other tasks involving quantum hypothesis testing are, e.g., information processing \cite{Hayashi06} and communications~\cite{holevo1973bounds,hayashi2003general,wangrenner2012,matthews2014finite}.% or signal processing~\cite{eldaropp2002quantum}.
When the number of hypotheses is two, quantum hypothesis testing allows a simple formulation in terms of two kinds of pairwise errors. The quantum version of the Neyman-Pearson lemma establishes the optimum binary test in this setting.
This problem was first studied by Helstrom in~\cite{helstrom1967detection} (see also \cite{bakut1968optimal, holevo1972decision}). When the number of hypotheses is larger than two, a (classical) prior distribution is usually placed over the hypotheses. While there exists no closed form for the optimal test in general, optimality contitions can be obtained~\cite{holevo1973decision,ykl1975optimum}. For historical notes on the subject see {\cite[Ch.~IV]{Helstrong76}}.

In the context of reliable communication, hypothesis testing has been instrumental in the derivation of converse bounds to the error probability both in the classical and quantum settings (see, e.g.,~\cite{Shan67,nagaoka2001strong}). Recently, hypothesis testing gained interest as a very general approach to obtain converse bounds in the finite block-length regime.
In classical channel coding, Polyanskiy, Poor and Verd{\'u} derived the meta-converse bound based on an instance of binary hypothesis testing~\cite{Pol10}. A similar approach was used by Wang and Renner to derive a finite block-length converse bound for classical-quantum channels~\cite{wangrenner2012}, and by Matthews and Wehner to obtain a family of converse bounds for general quantum channels~\cite{matthews2014finite}. The results by Matthews and Wehner are general enough to recover the meta-converse bound in the classical setting and Wang-Renner converse bound in the classical-quantum setting.

The information-spectrum method studies the asymptotics of a certain random variable, often referred to as information density or information random variable. Using a quantum analogue of this quantity,
Hayashi and Nagaoka studied quantum hypothesis testing~\cite{nagaoka2007ht}, and classical-quantum channel coding~\cite{hayashi2003general}, obtaining general bounds for both problems.

In this paper, we derive two alternative exact expressions for the minimum error probability of multiple quantum hypothesis testing when a (classical) prior distribution is placed over the hypotheses. The expressions obtained illustrate connections among hypothesis testing, information-spectrum measures and converse bounds in classical-quantum channel coding. An application to classical-quantum channel coding shows that Matthews-Wehner converse bound \cite[Th.~19]{matthews2014finite} and Hayashi-Nagaoka lemma\mbox{\cite[Lemma~4]{hayashi2003general}} with certain parameters yield the exact error probability. This work thus generalizes several results derived in~\cite{arxiv15ht} in the classical setting.

\section{Background}\label{sec:QHT}
%%%%%%%%%%%%%%%%%%%%%%%%%%%%%%%%%%%%%%%%%%%%%%%%%%%%%%%%%%%%%%%
%We first introduce some notation and well-known results.

\subsection{Notation}
In the general case, a quantum state is described by a density operator $\rho$ acting on some finite dimensional complex Hilbert space~$\Hc$. Density operators are self-adjoint, positive semidefinite, and have unit trace.
A measurement on a quantum system is a mapping from the state of the system $\rho$ to a classical outcome $m\in\{1,\ldots,M \}$. A measurement is represented by a collection of positive self-adjoint operators $\bigl\{\Pi_1,\ldots,\Pi_M\bigr\}$ such that $\sum \Pi_m = \openone$, where $\openone$ is the identity operator. These operators form a POVM (positive operator-valued measure). A measurement $\bigl\{\Pi_1,\ldots,\Pi_M\bigr\}$ applied to $\rho$ has outcome $m$ with probability $\tr(\rho\Pi_m)$. %The state after the measurement, conditioned on the outcome $m$, is $\Pi_m^{\frac{1}{2}} \rho \Pi_m^{\frac{1}{2}} / \tr\bigl(\Pi_m \rho\bigr)$.

For two self-adjoint operators $A,B$, the notation $A \geq B$ means that $A - B$ is positive semidefinite. Similarly $A \leq B$, $A > B$, and $A < B$ means that $A - B$ is negative semidefinite, positive definite and negative definite, respectively. For a self-adjoint operator $A$ with spectral decomposition $A = \sum_i \lambda_i E_i$, where $\{\lambda_i\}$ are the eigenvalues and $\{E_i\}$ are the orthogonal projections onto the corresponding eigenspaces, we define
\begin{align}
  \{A > 0\} \triangleq \sum_{i:\lambda_i > 0} E_i.
\end{align}
This corresponds to the projector associated to the positive eigenspace of $A$. We shall also use $\{A \geq 0\} \triangleq \sum_{i:\lambda_i \geq 0} E_i$, $\{A < 0\} \triangleq \sum_{i:\lambda_i < 0} E_i$ and $\{A \leq 0\} \triangleq \sum_{i:\lambda_i \leq 0} E_i$.

\subsection{Binary Hypothesis Testing}\label{sec:bary}
%%%%%%%%%%%%%%%%%%%%%%%%%%%%%%%%%%%%%%%%%%%%%%%%%%%%%%%%%%%%%%%%%%%%%%%%%%%

Let us consider a binary hypothesis test (with simple hypotheses) discriminating between the density operators $\rho_0$ and $\rho_1$ acting on $\Hc$. In order to distinguish between the two hypotheses we perform a measurement. We define a test measurement $\{T, \bar T\}$, such that $T$ and $\bar T \triangleq \openone - T$ are positive semidefinite. The test decides $\rho_0$ (resp. $\rho_1$) when the measurement outcome corresponding to $T$ (resp. $\bar T$) occurs.

Let $\eps_{j|i}$ denote the probability of deciding $\rho_j$ when $\rho_i$ is the true hypothesis, $i,j=0,1$, $i\neq j$. More precisely, we define
\begin{align}
 \eps_{1|0}(T) &\triangleq  1- \tr\left(\rho_0 T\right) = \tr\left(\rho_0\bar T\right), \label{eqn:bht-pi10}\\
 \eps_{0|1}(T) &\triangleq \tr\left(\rho_1 T\right).     \label{eqn:bht-pi01}
\end{align}

Let $\alpha_{\beta}(\rho_0 \| \rho_1)$ denote the minimum error probability $\eps_{1|0}$ among all tests with $\eps_{0|1}$ at most $\beta$, that is,
\begin{align}
\alpha_{\beta}(\rho_0 \| \rho_1)
  \triangleq \inf_{T: \eps_{0|1}(T) \leq \beta}  \eps_{1|0}(T).
 \label{eqn:bht-alpha}
\end{align}
The function $\alpha_{\beta}(\cdot\|\cdot)$ is the inverse of the function $\beta_{\alpha}(\cdot\|\cdot)$ appearing in~\cite{matthews2014finite}, which is itself related to the hypothesis-testing relative entropy as $D_{\text{H}}^{\alpha}(\rho_0 \| \rho_1) = - \log \beta_{\alpha}(\rho_0 \| \rho_1)$~\cite{wangrenner2012}.

When $\rho_0$ and $\rho_1$ commute, the test $T$ in~\refE{bht-alpha} can be restricted to be diagonal in the (common) eigenbasis of $\rho_0$ and $\rho_1$, then \refE{bht-alpha} reduces to the classical case~\cite{arxiv15ht}.

The quantum version of the Neyman-Pearson lemma characterizes the form of the test  minimizing \refE{bht-alpha}.
Let $t \geq 0$ and let $P^{+}_{t}$, $P^{-}_{t}$, $P^{0}_{t}$ denote the projectors spanning the positive, negative and null eigenspaces of the matrix $\rho_0 - t \rho_1$, respectively, i.~e.,
\begin{align}
  P^{+}_{t} &\triangleq \bigl\{\rho_0 - t \rho_1 > 0\bigr\},\\
  P^{-}_{t} &\triangleq \bigl\{\rho_0 - t \rho_1 < 0\bigr\},\\
  P^{0}_{t} \,&\triangleq\, \openone - P^{+}_{t} - P^{-}_{t}.
\end{align}

\begin{lemma}[Neyman-Pearson lemma]\label{lem:NPlemma}
The operator $T_{\text{NP}}$ is an optimal test between $\rho_0$ and $\rho_1$ if and only if
\begin{align}
  T_{\text{NP}} = P^{+}_{t} + p^{0}_{t},
\end{align}
where $0 \leq p^{0}_{t} \leq P^{0}_{t}$.
\end{lemma}
\begin{IEEEproof}
A slightly different formulation of this result is usually given in the literature. The statement included here can be found in, e.g., \cite[Lem. 3]{jenvcova2010quantum}. 
\end{IEEEproof}

Therefore, for any $t \geq 0$ and $0 \leq p^{0}_{t} \leq P^{0}_{t}$ such that $\tr\bigl\{\rho_1 T_{\text{NP}}\bigr\} = \beta$, the resulting test $T_{\text{NP}}$ minimizes~\refE{bht-alpha}.
%This fact shall be used to evaluate the function $\alpha_{\beta}(\cdot\|\cdot)$.
Moreover, the following lower bound holds.
\begin{lemma}\label{lem:NPbound}
For any test discriminating between $\rho_0$ and $\rho_1$, and for any $t' \geq 0$, it holds that
\begin{align}
\alpha_{\beta}(\rho_0 \| \rho_1)
\,\geq\, \tr\Bigl( \rho_0 \bigl(P^{-}_{t'}+P^{0}_{t'}\bigr) \Bigr)  - t' \beta.
\end{align}
\end{lemma}
\begin{IEEEproof}
For any operator $A\geq 0$  and $0 \leq T \leq \openone$,
it holds that $\tr\bigl( A \{ A > 0\} \bigr) \geq \tr\bigl( A T \bigr)$~\cite[Eq. 8]{nagaoka2007ht}. For $A = \rho_0-t'\rho_1$ and $T=T_{\text{NP}}$, this inequality becomes
\begin{align}
  \tr\bigl( (\rho_0-t'\rho_1) P^{+}_{t'} \bigr)
  \geq \tr\bigl( (\rho_0-t'\rho_1) T_{\text{NP}} \bigr),
  \label{eqn:NPbound-1}
\end{align}
which after some algebra yields
\begin{align}
  -\!\tr\bigl( \rho_0 T_{\text{NP}} \bigr)
       \geq -\!\tr\bigl( \rho_0 P^{+}_{t'} \bigr)
            + t' \tr\bigl(\rho_1 (P^{+}_{t'}-T_{\text{NP}}) \bigr).\label{eqn:NPbound-2}
\end{align}
Summing one to both sides of \refE{NPbound-2} and noting that $\alpha_{\beta}(\rho_0 \| \rho_1) = 1-\tr\bigl( \rho_0 T_{\text{NP}}\bigr)$ and $\beta = \tr\bigl(\rho_1 T_{\text{NP}} \bigr)$, we obtain
\begin{multline}
\alpha_{\beta}(\rho_0 \| \rho_1)
   \geq \tr\bigl( \rho_0 (P^{-}_{t'}+P^{0}_{t'}) \bigr)
           \!+t'\!\tr\bigl(\rho_1 P^{+}_{t'}\bigr)\!-t'\beta.\label{eqn:NPbound-3}
\end{multline}
The result thus follows by lower-bounding $\tr\bigl(\rho_1 P^{+}_{t'}\bigr) \geq 0$.
\end{IEEEproof}

\section{Multiple Quantum Hypothesis Testing}\label{sec:mary}
%%%%%%%%%%%%%%%%%%%%%%%%%%%%%%%%%%%%%%%%%%%%%%%%%%%%%%%%%%%%%%%%%%%%%%%%%%%

We consider a hypothesis testing problem discriminating among $M$ possible states acting on $\Hc$, where $M$ is assumed to be finite. The $M$ alternatives $\tau_1, \ldots, \tau_M$ are assumed to occur with (classical) probabilities $p_1, \ldots, p_M$, respectively.

A $M$-ary hypothesis test is a POVM $\Pc\!\triangleq\!\{ \Pi_{1},  \Pi_{2}, \ldots, \Pi_{M}\!\}$, $\sum \Pi_i = \openone$.
The test decides the alternative $\tau_i$ when the measurement with respect to $\Pc$ has outcome $i$.
The probability that the test $\Pc$ decides $\tau_j$ when $\tau_i$ is the true underlying state is thus $\tr \bigl(\tau_i \Pi_{j} \bigr)$ and the average error probability is
\begin{align}
  \epsilon(\Pc) \triangleq  1 - \sum_{i=1}^M p_i \tr\left(\tau_i \Pi_i\right).
  \label{eqn:mht-eps}
\end{align}
We define the minimum average error probability as
\begin{align}
  \epsilon \triangleq \min_{\Pc} \epsilon(\Pc).
  \label{eqn:mht-mineps}
\end{align}
The test $\Pc$ minimizing \refE{mht-mineps} has no simple form in general.

\begin{lemma}[Holevo-Yuen-Kennedy-Lax conditions]\label{lem:Pstar}A test $\Pc^{\star} = \{ \Pi_{1}^{\star},  \ldots, \Pi_{M}^{\star} \}$ minimizes~\refE{mht-mineps} if and only if, for each $m=1,\ldots,M$,
\begin{align}
  \bigl( \Lambda(\Pc^{\star}) - p_m \tau_m \bigr)\Pi_m^{\star} 
    \,=\, \Pi_m^{\star}\bigl( \Lambda(\Pc^{\star}) - p_m \tau_m \bigr) &\,=\, 0,
    \label{eqn:mht-pistar1}\\
  \Lambda(\Pc^{\star}) - p_m \tau_m &\,\geq\, 0,
    \label{eqn:mht-pistar2}
\end{align}
where 
\begin{align}\label{eqn:Upsilon-def}
       \Lambda(\Pc^{\star}) &\triangleq \sum_{i=1}^{M} p_i \tau_i \Pi_i^{\star} = \sum_{i=1}^{M} p_i \Pi_i^{\star} \tau_i
\end{align}
is required to be self-adjoint\footnote{The operator $\Lambda(\Pc)$ takes a role of the Lagrange multiplier associated to the constraint $\sum \Pi_{i} = \openone$, which, involving self-adjoint operators requires $\Lambda$ to be self-adjoint.}.
\end{lemma}
\begin{IEEEproof}
The theorem follows from \cite[Th. 4.1, Eq. (4.8)]{holevo1973decision} or \cite[Th. I]{ykl1975optimum} after simplifying the resulting optimality conditions.
 % \cite[Sec. 2]{belavkin1975optimal}.
\end{IEEEproof}

%From the definition of $\Lambda(\cdot)$ in \refE{Upsilon-def}, it follows that
%\begin{align}
%  \tr\bigl(\Lambda(\Pc)\bigr)
%    &= \frac{1}{M} \sum_{i=1}^{M} \sum_{j\neq i} \tr\bigl(\Pi_i \tau_j\bigr)\\
%    &= \frac{1}{M} \sum_{j=1}^{M} \tr\Biggl( \Biggl(\sum_{i\neq j} \Pi_i \Biggr)\tau_j \Biggr)\\
%    &= 1 - \frac{1}{M} \sum_{j=1}^{M} \tr\bigl(\Pi_j \tau_j\bigr)\\
%    &= \epsilon(\Pc),
%\end{align}
%and, as a result, $\epsilon = \tr\bigl(\Lambda(\Pc^{\star})\bigr)$.

%\subsection{Main Result}
%%%%%%%%%%%%%%%%%%%%%%%%%%%%%%%%%%%%%%%%%%%%%%%
We next show an alternative characterization of the minimum error probability $\epsilon$ as a function of a binary hypothesis test with certain parameters.

Let $\diag(\rho_1, \ldots, \rho_M)$ denote the block-diagonal matrix with diagonal blocks $\rho_1, \ldots, \rho_M$. We define \begin{align}
\Tc &\triangleq \diag \bigl(p_1 \tau_1, \ldots, p_M \tau_M\bigr),
\label{eqn:Tc-def}\\
\Dc(\mu_0) &\triangleq \diag \bigl(\tfrac{1}{M} \mu_0, \ldots, \tfrac{1}{M}\mu_0\bigr),
\label{eqn:Dc-def}
\end{align}
where $\mu_0$ is an arbitrary density operator acting on $\Hc$. Note that both $\Tc$ and $\Dc(\mu_0)$ are density operators themselves, % acting on $\Hc\otimes\CC^M$,
since they are self-adjoint, positive semidefinite and have unit trace.

\begin{theorem}\label{thm:main-result} %%%%%%%%%%%%%%%%%%%%%%%%%%%%%%%%%
The minimum error probability of an \mbox{$M$-ary} test discriminating among
states $\{\tau_1,\ldots,\tau_M\}$ with prior classical probabilities $\{p_1, \ldots, p_M\}$ satisfies
\begin{align}
  \epsilon = \max_{\mu_0} \alpha_{\frac{1}{M}} \bigl(\Tc \,\|\, \Dc(\mu_0)\bigr),
  \label{eqn:main-result}
\end{align}
where $\Tc$ and $\Dc(\cdot)$ are given in \refE{Tc-def} and \refE{Dc-def}, respectively, and where the optimization is carried out over (unit-trace non-negative) density operators $\mu_0$.
\end{theorem} %%%%%%%%%%%%%%%%%%%%%%%%%%%%%%%%%%%%%%%%%%%%%%%%%%%%%%%%%%%%%%%%%%
\begin{IEEEproof}
For any $\Pc = \{ \Pi_{1},  \Pi_{2}, \ldots, \Pi_{M} \}$ let us define the binary test $T' \triangleq \diag \left(\Pi_1, \ldots, \Pi_M\right)$. For this test we obtain
\begin{align}
 \eps_{1|0}(T') &=  1 - \sum_{i=1}^M p_i \tr\left(\tau_i \Pi_i\right) = \epsilon(\Pc),
                \label{eqn:pi10-Pitilde}\\
 \eps_{0|1}(T') &= \frac{1}{M} \sum_{i=1}^M \tr\left(\mu_0 \Pi_i\right)\\
                 &= \frac{1}{M} \tr\left(\mu_0 \left(\sum\nolimits_{i=1}^M \Pi_i\right) \right)\\
                 &= \frac{1}{M} \tr\left( \mu_0 \right) = \frac{1}{M}.\label{eqn:pi01-Pitilde}
\end{align}
The (possibly suboptimal) test $T'$ has thus $\eps_{1|0}(T') = \epsilon(\Pc)$ and $\eps_{0|1}(T') = \frac{1}{M}$. Therefore, using \refE{bht-alpha} and maximizing the resulting expression over $\mu_0$, we obtain
\begin{align}
  \epsilon(\Pc) \geq  \max_{\mu_0} \alpha_{\frac{1}{M}} \bigl(\Tc\,\|\, \Dc(\mu_0)\bigr).
  \label{eqn:lower-bound}
\end{align}

It remains to show that, for  $\Pc = \Pc^{\star}$ defined in Lemma~\ref{lem:Pstar}, the lower bound \refE{lower-bound} holds with equality.
To this end, we next demonstrate that the optimality conditions for $T_{\text{NP}}$ in Lemma~\ref{lem:NPlemma} and for $\Pc^{\star} = \{ \Pi_{1}^{\star},  \ldots, \Pi_{M}^{\star} \}$ in Lemma~\ref{lem:Pstar} are equivalent for a specific choice of $\mu_0$.

Let $\Pc^{\star} = \{ \Pi_{1}^{\star},  \ldots, \Pi_{M}^{\star} \}$ satisfy \refE{mht-pistar1}-\refE{mht-pistar2} and define
\begin{align}\label{eqn:mu0star}
   \mu_0^{\star} \triangleq \frac{1}{c_0^{\star}} \sum_{i=1}^M p_i \tau_i \Pi_{i}^{\star} = \frac{1}{c_0^{\star}}  \Lambda(\Pc^{\star}),
\end{align}
where $c_0^{\star}$ is a normalizing constant such that $\mu_0^{\star}$ is unit trace. 

Lemma \ref{lem:NPlemma} shows that the test $T_{\text{NP}}$ achieving \refE{lower-bound} is associated to the non-negative eigenspace of the matrix $\Tc-t \Dc(\mu_0)$. Given the block-diagonal structure of the matrix $\Tc-t \Dc(\mu_0)$, it is enough to consider binary tests $T_{\text{NP}}$ with block-diagonal structure. Then, we write $T_{\text{NP}} = \diag \left(T_1^{\text{NP}}, \ldots, T_M^{\text{NP}}\right)$.

For the choice $\mu_0 = \mu_0^{\star}$, and $t = M c_0^{\star}$, the $m$-th block-diagonal term in $\Tc-t \Dc(\mu_0)$ is given by
\begin{align}
   p_m\tau_m - \tfrac{t}{M} \mu_0
     &= p_m\tau_m - \Lambda(\Pc^{\star}).
     \label{eqn:blockTlambdaD-1}
\end{align}

The $m$-th block of the Neyman-Pearson test $T_m^{\text{NP}}$ must lie in the non-negative eigenspace of the matrix \refE{blockTlambdaD-1}. However, since \refE{mht-pistar2} implies that \refE{blockTlambdaD-1} is negative semidefinite, each block $T_m^{\text{NP}}$ can only lie in the null eigenspace of~\refE{blockTlambdaD-1}, $m=1,\ldots,M$.
%Using the notation in Lemma~\ref{lem:NPlemma}, thus $P^{+}_{t} = 0$ and $T_{\text{NP}} = p^{0}_{t}$.

According to \refE{mht-pistar1}, the operator $\Pi_{m}^{\star}$ belongs to the null eigenspace of \refE{blockTlambdaD-1}, $m=1,\ldots,M$. As a result, the choice 
\begin{align}
T_{\text{NP}}  
  &= \diag \left(\Pi_1^{\star}, \ldots, \Pi_M^{\star}\right)
\end{align}
satisfies the optimality conditions in Lemma~\ref{lem:NPlemma}. Moreover, since $\eps_{1|0}(T_{\text{NP}}) = \epsilon\bigl(\Pc^{\star}\bigr) = \epsilon$ and $\eps_{0|1}(T_{\text{NP}}) = \frac{1}{M}$,  Lemma~\ref{lem:NPlemma} implies that \refE{main-result} holds with equality for $\mu_0 = \mu_0^{\star}$. Given the bound in \refE{lower-bound}, other choices of $\mu_0$ cannot improve the result, and Theorem \ref{thm:main-result} thus follows.
\end{IEEEproof}

Combining Theorem~\ref{thm:main-result} and Lemma \ref{lem:NPbound}, we obtain a characterization for $\epsilon$ based on information-spectrum measures.

%The expression \refE{main-result} in Theorem \ref{thm:main-result} is no easier to compute that the minimization in \refE{mht-mineps}. To see this note that each of the blocks \refE{blockTlambdaD} depend on $\Lambda(\Pc^{\star})$ which itself depends on the optimal $M$-ary test $\Pc^{\star}$.

\begin{theorem}\label{thm:tight-spectrum} %%%%%%%%%%%
The minimum error probability of an \mbox{$M$-ary} test discriminating among states $\{\tau_1,\ldots,\tau_M\}$ with prior classical probabilities $\{p_1, \ldots, p_M\}$ satisfies
\begin{align}
  \epsilon = \max_{\mu_0, t\geq 0} \left\{ \sum_{i=1}^{M} p_i \tr\Bigl( \tau_i \bigl\{ p_i \tau_i - t \mu_0 \leq 0 \bigr\} \Bigr)- t \right\}\!.
\label{eqn:tight-spectrum}
\end{align}
where the optimization is carried out over (unit-trace non-negative) density operators $\mu_0$ acting on $\Hc$, and over the scalar threshold $t \geq 0$.
\end{theorem} %%%%%%%%%%%%%%%%%%%%%%%%%%%%%%%
\begin{IEEEproof}
Applying Lemma~\ref{lem:NPbound} to \refE{main-result}, and using 
%\begin{align}
%  \epsilon \,\geq\,
%    \tr\Bigl( \Tc \bigl\{ \Tc -t' \Dc(\mu_0) \leq 0 \bigr\} \Bigr) 
%            - \frac{t'}{M}.
%\label{eqn:lemma-hn-1}
%\end{align}
the definitions of $\Tc$ in \refE{Tc-def} and $\Dc(\cdot)$ in \refE{Dc-def}, yields, for any $\mu_0$, $t'\geq 0$,
\begin{align}
  \epsilon \geq \sum_{i=1}^{M} p_i \tr\Bigl( \tau_i \bigl\{ p_i \tau_i - \tfrac{t'}{M} \mu_0 \leq 0 \bigr\} \Bigr)- \tfrac{t'}{M}.
\label{eqn:lemma-hn-2}
\end{align}

It remains to show that there exist $\mu_0$ and $t'\geq 0$ such that \refE{lemma-hn-2} holds with equality.
In particular, let us choose $\mu_0 = \mu_0^{\star}$ defined in~\refE{mu0star}, and $t' = M c_0^{\star}$ where
%\begin{align}
   $c_0^{\star} = \sum_{i=1}^M p_i \tr(\tau_i \Pi_{i}^{\star})$
%\end{align}
is the normalizing constant from~\refE{mu0star}.

For this choice of $\mu_0$ and $t'$, the projector spanning the negative semidefinite eigenspace of the operator \mbox{$p_i\tau_i - \frac{t'}{M} \mu_0$} can be rewritten as
\begin{align}
 \Bigl\{ p_i \tau_i -\tfrac{t'}{M} \mu_0 \leq 0 \Bigr\}
&= \bigl\{ p_i \tau_i - \Lambda(\Pc^{\star}) \leq 0 \bigr\}\\
&= \openone, \label{eqn:trivialproj}
\end{align}
where the last identity follows from \refE{mht-pistar2}.
The right-hand side of \refE{lemma-hn-2} thus becomes
\begin{align}
\sum_{i=1}^{M} p_i \tr( \tau_i ) - \frac{t'}{M} 
\,=\, 1 - \frac{t'}{M}.
\label{eqn:lemma-hn-3}
\end{align}
The result follows since $\tfrac{t'}{M}\!=\! c_0^{\star}\!=\!\sum_{i} p_i \tr(\tau_i\Pi_{i}^{\star})\!=\!1\!-\!\eps$.
\end{IEEEproof}

The alternative expressions derived in Theorems \ref{thm:main-result} and \ref{thm:tight-spectrum} are not easier to compute than the original optimization in \refE{mht-mineps}, all of them requiring to solve a semidefinite program.
We recall from the proofs of the theorems that a density operator $\mu_0$ maximizing \refE{main-result} and \refE{tight-spectrum} is 
\begin{align}
   \mu_0^{\star} = \frac{1}{c_0^{\star}} \sum_{i=1}^M p_i \tau_i \Pi_{i}^{\star},
\end{align}
for some $\Pc^{\star} = \{ \Pi_{1}^{\star},  \ldots, \Pi_{M}^{\star} \}$ satisfying the conditions in Lemma \ref{lem:Pstar} and where $c_0^{\star}$ is a normalizing constant. Hence, the optimal $M$-ary hypothesis test $\Pc^{\star}$ characterizes the optimal $\mu_0$. Conversely, the optimal $\mu_0$ is precisely the Lagrange multiplier associated to the minimization in \refE{mht-mineps}, after an appropriate re-scaling.

The expressions derived in Theorems \ref{thm:main-result} and \ref{thm:tight-spectrum} can be used to determine the tightness of several converse bounds from the literature, as we show in the next section.

\section{Application to Classical-Quantum Channels}\label{sec:cc}
%%%%%%%%%%%%%%%%%%%%%%%%%%%%%%%%%%%%%%%%%%%%%%%%%%%%%%%%%%%%%%%%%%%%%%%%%%
We consider the channel coding problem of transmitting  $M$ equiprobable messages over a one-shot classical-quantum channel $x \to W_x$, with $x\in\Xc$ and $W_x\in\Hc$.
% We define the one-shot classical-quantum channel with input alphabet $\Xc=\{1,\ldots,|\Xc|\}$ and associated density operators $W_x$, $x\in\Xc$, in a finite dimensional Hilbert space $\Hc$.

A channel code is defined as a mapping from the message set $\{1,\ldots,M\}$ into a set of $M$ codewords $\Cc = \{x_1,\ldots,x_M\}$. 
For a source message $m$, the decoder receives the associated density operator $W_{x_m}$ and must decide on the transmitted message. The minimum error probability for a code $\Cc$ is
\begin{align}
  \Pe(\Cc) 
    &\triangleq  \min_{\{\Pi_1,\ldots,\Pi_M\}} \left\{ 1 - \frac{1}{M} \sum_{m=1}^M \tr\bigl(W_{x_m} \Pi_m\bigr) \right \}.
\end{align}
This problem corresponds precisely to the $M$-ary quantum hypothesis testing problem described in Section \ref{sec:mary}.
Then, direct application of Theorems \ref{thm:main-result} and~\ref{thm:tight-spectrum} yields two alternative expressions for $\Pe(\Cc)$.

Let $\AA$ and $\BB$ denote the input and output of the system, respectively. The joint state induced by a codebook $\Cc$ is
\begin{align}
  \rho^{\AA\BB}_{\Cc} = \frac{1}{M} \sum_{x\in\Cc} \left| x \rangle \langle x \right|^{\AA} \otimes W_x^{\BB},
\end{align}
and $\rho^{\AA}_{\Cc} = \frac{1}{M} \sum_{x\in\Cc} \left| x \rangle \langle x \right|^{\AA}$ its input marginal.

According to \refE{main-result} in Theorem \ref{thm:main-result} we obtain
\begin{align}
  \Pe(\Cc) &= \max_{\mu_0} \alpha_{\frac{1}{M}} \bigl(\rho^{\AA\BB}_{\Cc} \,\|\, \rho^{\AA}_{\Cc} \otimes \mu_0^{\BB}\bigr).
    \label{eqn:alpha-cc}
\end{align}

The expression \refE{alpha-cc} is precisely the finite block-length converse bound by Matthews and Wehner \cite[Eq.~(45)]{matthews2014finite}, particularized for a classical-quantum channel with an input state induced by the codebook $\Cc$. Therefore, Theorem \ref{thm:main-result} implies that the quantum generalization of the meta-converse bound proposed by Matthews and Wehner is tight for a fixed codebook $\Cc$. % This fact was also observed for the classical meta-converse bound~\cite{arxiv15ht}.

Minimizing the right-hand side of \refE{alpha-cc} over all distributions $\px$ defined over the input alphabet $\Xc$, not necessarily induced by a codebook, yields a lower bound on $\Pe(\Cc)$ for any codebook $\Cc$. By fixing $\mu_0$ to be the state induced at the system output, this lower bound recovers the converse bound by Wang and Renner \cite[Th.~1]{wangrenner2012}. This bound is not tight in general since (i) the minimizing $\px$ does not need to coincide with the input state induced by the best codebook, and (ii) the choice of $\mu_0$ in \cite[Th. 1]{wangrenner2012} does not maximize the resulting bound in general.

Using the characterization in Theorem~\ref{thm:tight-spectrum}, the error probability $\Pe(\Cc)$ can be equivalently written as
\begin{align}
  \Pe(\Cc)\!=\! \max_{\mu_0, t'\geq 0} \left\{ \frac{1}{M} \sum_{x\in\Cc}\tr\Bigl( W_x \bigl\{ W_x\!-\!t'\mu_0 \leq 0 \bigr\} \Bigr)- \frac{t'}{M} \right\}\!.
\label{eqn:tight-hn}
\end{align}
The objective of the maximization in~\refE{tight-hn} coincides with the information-spectrum bound \cite[Lemma~4]{hayashi2003general}. Then, \refE{tight-hn} shows that the Hayashi-Nagaoka lemma yields the exact error probability for a fixed code, after optimizantion over the free parameters $\mu_0$, $t'\geq 0$.

\section{Concluding Remarks}\label{sec:discussion}
%%%%%%%%%%%%%%%%%%%%%%%%%%%%%%%%%%%%%%%%%%%%%%%%%%%%%%%%%%%%%%%

In Theorem \ref{thm:main-result}, the minimum error probability of an \mbox{$M$-ary} quantum hypothesis test is expressed as an instance of a binary quantum hypothesis test with certain parameters. This expression implies the tightness of the converse bound\mbox{\cite[Th.~19]{matthews2014finite}} by Matthews and Wehner, and identifies the weakness of \cite[Th. 1]{wangrenner2012} by Wang and Renner in classical-quantum channel coding. For more general channels and entanglement-assisted codes, it is not clear whether the bounds in~\mbox{\cite[Th.~18~and~Th.~19]{matthews2014finite}} coincide with the exact error probability. To study this, a generalization of Theorem~\ref{thm:main-result} imposing less structure over the test alternatives is needed.
Theorem~\ref{thm:tight-spectrum} shows that the minimum error probability can be written as an optimization problem involving information-spectrum measures. In particular, this expression shows that the Hayashi-Nagaoka lemma~\cite[Lemma~4]{hayashi2003general} yields the exact error probability after optimizantion over its free parameters.

\section*{Acknowledgment}
The problem studied here was suggested to the author by Alfonso Martinez. The author thanks him, Albert Guill\`en i F\`abregas and William Matthews for stimulating discussions related to this work.

\bibliographystyle{IEEEtran}
\bibliography{bib/references}

\begin{thebibliography}{10}
\providecommand{\url}[1]{#1}
\csname url@samestyle\endcsname
\providecommand{\newblock}{\relax}
\providecommand{\bibinfo}[2]{#2}
\providecommand{\BIBentrySTDinterwordspacing}{\spaceskip=0pt\relax}
\providecommand{\BIBentryALTinterwordstretchfactor}{4}
\providecommand{\BIBentryALTinterwordspacing}{\spaceskip=\fontdimen2\font plus
\BIBentryALTinterwordstretchfactor\fontdimen3\font minus
  \fontdimen4\font\relax}
\providecommand{\BIBforeignlanguage}[2]{{%
\expandafter\ifx\csname l@#1\endcsname\relax
\typeout{** WARNING: IEEEtran.bst: No hyphenation pattern has been}%
\typeout{** loaded for the language `#1'. Using the pattern for}%
\typeout{** the default language instead.}%
\else
\language=\csname l@#1\endcsname
\fi
#2}}
\providecommand{\BIBdecl}{\relax}
\BIBdecl

\bibitem{helstrom1967detection}
C.~W. Helstrom, ``Detection theory and quantum mechanics,'' \emph{Inf. and
  Control}, vol.~10, no.~3, pp. 254--291, 1967.

\bibitem{bakut1968optimal}
P.~A. Bakut and S.~S. Shchurov, ``Optimal detection of a quantum signal,''
  \emph{Probl. Peredachi Inf.}, vol.~4, no.~1, pp. 77--82, 1968, (in Russian,
  English translation: {\it Probl. Inf. Transm.}, vol. 4, pp. 61{-–}65,
  1968).

\bibitem{holevo1972decision}
A.~S. Holevo, ``An analog of the theory of statistical decisions in
  noncommutative theory of probability,'' \emph{Trudy Moskov. Mat.
  Ob\v{s}\v{c}.}, vol.~26, pp. 133--149, 1972, (in Russian).

\bibitem{holevo1973decision}
------, ``Statistical decision theory for quantum systems,'' \emph{J.
  Multivariate Anal. 3}, vol.~3, no.~4, pp. 337--394, 1973.

\bibitem{ykl1975optimum}
H.~P. Yuen, R.~S. Kennedy, and M.~Lax, ``Optimum testing of multiple hypotheses
  in quantum detection theory,'' \emph{IEEE Trans. Inf. Theory}, vol.~21,
  no.~2, pp. 125--134, Mar 1975.

\bibitem{Helstrong76}
C.~W. Helstrom, \emph{Quantum Detection and Estimation Theory}.\hskip 1em plus
  0.5em minus 0.4em\relax NY: Academic Press, 1976.

\bibitem{Shan67}
C.~E. Shannon, R.~G. Gallager, and E.~R. Berlekamp, ``{Lower bounds to error
  probability for coding on discrete memoryless channels. I},'' \emph{Inf.
  Contr.}, vol.~10, no.~1, pp. 65--103, 1967.

\bibitem{nagaoka2001strong}
H.~Nagaoka, ``Strong converse theorems in quantum information theory,'' in
  \emph{Proc. ERATO Conf. Quantum Inf. Science}, Tokyo, Japan, 2001, p.~33.

\bibitem{Pol10}
Y.~Polyanskiy, H.~V. Poor, and S.~Verd{\'u}, ``{Channel coding rate in the
  finite blocklength regime},'' \emph{IEEE Trans. Inf. Theory}, vol.~56, no.~5,
  pp. 2307--2359, 2010.

\bibitem{wangrenner2012}
L.~Wang and R.~Renner, ``One-shot classical-quantum capacity and hypothesis
  testing,'' \emph{Phys. Rev. Lett.}, vol. 108, no.~20, p. 200501, 2012.

\bibitem{matthews2014finite}
W.~Matthews and S.~Wehner, ``Finite blocklength converse bounds for quantum
  channels,'' \emph{IEEE Trans. Inf. Theory}, vol.~60, no.~11, pp. 7317--7329,
  2014.

\bibitem{nagaoka2007ht}
H.~Nagaoka and M.~Hayashi, ``An information-spectrum approach to classical and
  quantum hypothesis testing for simple hypotheses,'' \emph{IEEE Trans. Inf.
  Theory}, vol.~53, no.~2, pp. 534--549, 2007.

\bibitem{hayashi2003general}
M.~Hayashi and H.~Nagaoka, ``General formulas for capacity of classical-quantum
  channels,'' \emph{IEEE Trans. Inf. Theory}, vol.~49, no.~7, pp. 1753--1768,
  2003.

\bibitem{arxiv15ht}
G.~Vazquez-Vilar, A.~Tauste~Campo, A.~Guill\'en~i F\`abregas, and A.~Martinez,
  ``Bayesian {$M$}-ary hypothesis testing: The meta-converse and
  {V}erd\'u-{H}an bounds are tight,'' \emph{IEEE Trans. Inf. Theory}, 2016, to
  appear. Preprint available at arXiv:1411.3292.

\bibitem{jenvcova2010quantum}
A.~Jen\v{c}ov\'a, ``Quantum hypothesis testing and sufficient subalgebras,''
  \emph{Lett. Math. Phys.}, vol.~93, no.~1, pp. 15--27, 2010.

\end{thebibliography}

\end{document}